\documentclass[a4paper]{article}
\usepackage{acronym}
\usepackage{stfloats}

\renewcommand{\footnotesize}{\scriptsize}

\usepackage{enumitem}
\usepackage{mathtools}
\usepackage{siunitx}
\usepackage{xspace}
\usepackage{url}        
\usepackage{hyperref}   
\usepackage{amsmath}    
\usepackage{authblk} 
\usepackage{comment} 
\DeclareUnicodeCharacter{000B}{\relax}

\title{Leveraging Audio Representations for Vibration-Based Crowd Monitoring in Stadiums}

\author[1]{Yen Cheng Chang}
\author[1]{Jesse Codling}
\author[2]{Yiwen Dong}
\author[1]{Jiale Zhang}
\author[1]{Jiasi Chen}
\author[2]{Hae Young Noh}
\author[1]{Pei Zhang}

\affil[1]{University of Michigan, 1301 Beal Ave, Ann Arbor, Michigan 48105, USA}
\affil[2]{Stanford University, Stanford, California, USA}
\date{}  

\begin{document}

\maketitle
\begin{abstract}
Crowd monitoring in sports stadiums is important to enhance public safety and improve audience experience. 
Existing approaches mainly rely on cameras and microphones, which can cause significant disturbances and often raise privacy concerns. 
In this paper, we sense floor vibration, which provides a less disruptive and more non-intrusive way of crowd sensing, to predict crowd behavior. 
However, since the vibration-based crowd monitoring approach is newly developed, one main challenge is the lack of training data due to sports stadiums are usually large public spaces with complex physical activities.

In this paper, we present \emph{\textbf{Vi}bration \textbf{L}everage \textbf{A}udio (ViLA)}, a vibration-based method that reduces the dependency on labeled data by pre-training with unlabeled cross-modality data. 
\emph{ViLA} is first pretrained on audio data in an unsupervised manner and then fine-tuned with a minimal amount of in-domain vibration data.
By leveraging publicly available audio datasets, \emph{ViLA} learns the wave behaviors from audio and then adapts the representation to vibration, reducing the reliance on domain-specific vibration data.
Our real-world experiments demonstrate that pre-training the vibration model using publicly available audio data (YouTube8M) achieved up to a 5.8$\times$ error reduction compared to the model without audio pre-training.

\end{abstract}

\begin{figure}[h]
    \centering
    \includegraphics[width=1\linewidth ]{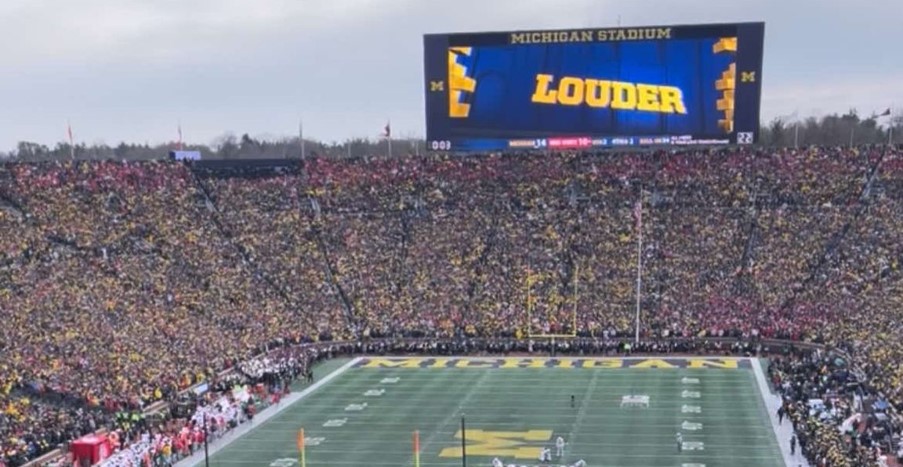}
    \caption{An overview of the football stadium. The complex crowd dynamics and high noise levels make crowd monitoring a challenging task.}
    \label{fig:umich_stadium}
\end{figure}

\section{Introduction}
Effective crowd monitoring enhances safety, security, and overall audience enjoyment at large-scale public events such as sports games \cite{zeitz2009crowd, filingeri2017factors, kingshott2014crowd}. 
Traditional methods like manual observation \cite{bahmanyar2019mrcnet} and the deployment of various sensing systems \cite{maheshwari2016review, jarvis2019miamimapper, yamin2018managing, zhang2025wevibeweightchangeestimation} have been successful in monitoring audience activities.
However, they face challenges such as requiring intensive manual labor, visual occlusions (computer vision), and insufficient accuracy due to time-varying noises from large groups (audio and WiFi/radio frequency) \cite{chang2024lts, dong2024ca, dong2024ubiquitous}.
In recent years, floor vibration sensing has emerged as a cost-efficient, non-intrusive, and privacy-friendly alternative, facilitating continuous and fine-grained monitoring of crowd behaviors \cite{pan2015indoor, codling2024flour, pei2020idiot}.
This new approach detects and analyzes crowed-induced vibrations without capturing an individual's voice or appearance, offering detailed insights into crowd traffic and movement patterns while operating unobtrusively.

However, collecting high-quality, well-labeled floor vibration data presents significant challenges due to the crowded environment at sports stadiums. 
First, deploying and synchronizing numerous sensors in large-scale, crowded spaces is time-consuming and complicated \cite{davies1995crowd, wirz2012inferring, jiang2021ultra}. 
Accurately calibrating and positioning these sensors is essential but challenging due to heavy foot traffic and interference from environmental noise \cite{pan2017, yves2017, majd2006, kessler2019, dong2023}. 
Additionally, floor vibrations are influenced by various factors such as material, weight, and location, which complicates the training of a general-purpose model that can cover all floor types and locations \cite{ljunggren2006floor, toratti2006classification, pan2017}.
Figure \ref{fig:umich_stadium} illustrates an example of large crowd monitoring. 
Thus, collecting data in such environments requires deploying numerous sensors, which face significant connectivity challenges in crowded settings\cite{dong2023}.
To the best of our knowledge, there is a lack of well-labeled datasets to allow efficient training of vibration-based models, which mainly use self-collected, small-scale datasets and focus on a single person's behaviors \cite{pan2017, dong2023characterizing, dong2024ubiquitous, dong2024ambient}.
Besides, some popular supervised techniques like transfer learning and domain adaptation can utilize these data for crowd monitoring, but they still require substantial labeled data for effective pre-training \cite{chang2023, sarkar2024, liu2023}.

\begin{figure*}[t]
    \centering
    \includegraphics[width=\linewidth]{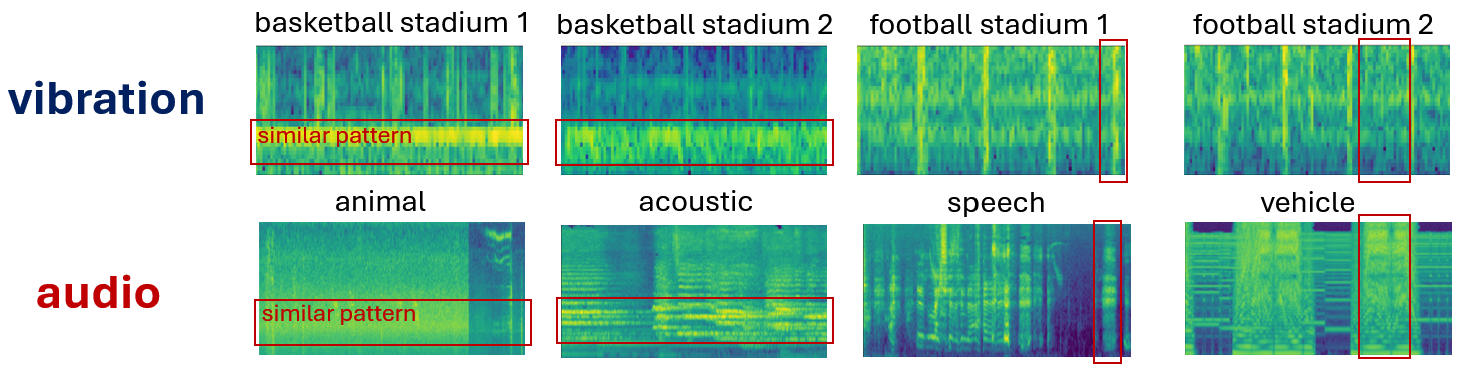}
    \caption{Examples of audio and vibration spectrograms. The horizontal axis represents time (seconds), and the vertical axis represents frequency. Both spectrograms exhibit many similar patterns.}
    \label{fig:audio_pretrain}
\end{figure*}

In this study, we propose \emph{ViLA}, a novel approach that leverages out-of-domain yet widely available datasets to pre-train the vibration-based crowd monitoring model. 
To compensate for the insufficient amount of vibration data, our approach leverages the abundant data resource from the chosen modality (i.e., audio) to capture the shared features in their time-frequency domain. 
Specifically, we first pre-train the crowd monitoring model using a publicly available audio dataset to capture basic features shared with the vibration data. 
Figure \ref{fig:audio_pretrain} shows some visual similarity between audio and vibration.
Then, we fine-tune the model using a small amount of labeled vibration data for crowd monitoring.  
The main benefit of our approach is that we reduce the reliance on large-scale labeled training data, which significantly lessens the manual effort for sensor deployment, data collection, and labeling.
The real-world evaluation at sports stadiums shows that using publicly available audio data (YouTube8M) for pre-training achieved up to a 5.8$\times$ reduction in error for vibration-based crowd behavior monitoring compared to the model without audio pre-training.

This paper introduces a novel approach to pre-train a vibration-based crowd monitoring model using a publicly available audio dataset. The main contributions of this paper are:
\begin{description}
    \item[(1)] Proposed Framework: We propose a \emph{ViLA}, a framework that leverages out-of-domain modality (i.e., audio) to improve the performance of vibration-based models for crowd monitoring.
    \item[(2)] Definitions of the two indicators: We introduce two indicators, \textit{Similarity} and \textit{Diversity}, to evaluate the selection of modalities from various publicly available datasets to effectively enhance the performance of vibration-based models.
    \item[(3)] Real-World Evaluation: We demonstrate the effectiveness of our approach through real-world evaluations at two sports stadiums, which show promising results.
\end{description}

We evaluate our approach by 1) analyzing various publicly available datasets, including audio, video, and image, and 2) conducting real-world experiments across two sports stadiums to collect crowd-induced floor vibration data. Our results show that audio signals enhance vibration-based models more effectively than other modalities using the \textit{Similarity} and \textit{Diversity} indicators. The real-world evaluation at sports stadiums shows that using publicly available audio data (YouTube8M) for pre-training achieved up to a 5.8$\times$ reduction in error for vibration-based crowd behavior monitoring compared to the model without audio pre-training. This demonstrates the effectiveness of our approach for efficient vibration-based model training and accurate crowd monitoring at sports stadiums.

The rest of the paper is organized as follows: we demonstrate how we select audio as our pre-training modality (Section 2). Then, we illustrate the framework that leverages unlabeled, pre-existing audio datasets to enhance vibration-based models (Section 3). Subsequently, we conduct real-world experiments in two different environments (Section 4). Finally, we summarize the study (Section 5).

\section{Unsupervised Pre-training Using Audio Data for Vibration Modeling} \label{sec:training_process}

In this section, we introduce a novel framework for vibration-based crowd monitoring model training that significantly reduces the reliance on labeled vibration data. Our approach consists of two steps: 1) unsupervised pre-training using public audio data and 2) supervised fine-tuning using minimal labeled vibration data specific to the target domain. 

This approach effectively addresses challenges such as data scarcity in harsh environments and reduces the effort and costs associated with data collection. By employing cross-domain learning techniques, our method not only bridges the gap between audio and vibration data but also enhances the efficiency and scalability of vibration monitoring systems.

\begin{figure*}[htb]
    \centering
    \includegraphics[width=\linewidth]{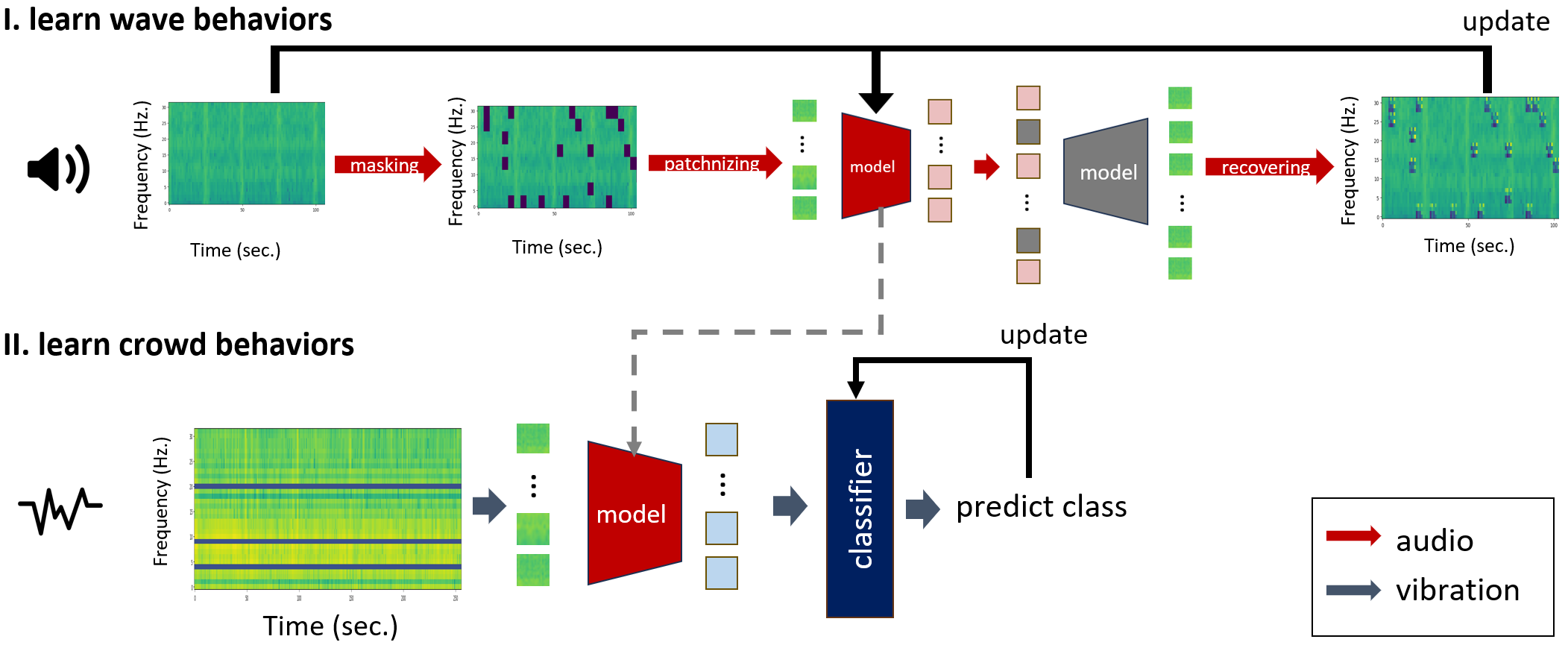}
    \caption{The overview of the \emph{ViLA} training process. Initially, unsupervised audio pre-training is performed. This is followed by supervised vibration fine-tuning using the encoder learned from the pre-training stage.}
    \label{fig:sec:training_process}
\end{figure*}

\subsection{The overview of the training process} 


The proposed method involves a two-stage training process, as illustrated in Figure \ref{fig:sec:training_process}. Initially, unsupervised pre-training utilizes the YouTube-8M dataset to train the model without requiring labeled vibration data. This stage focuses on capturing general features from diverse audio content. Subsequently, the model undergoes supervised fine-tuning using domain-specific vibration data obtained from environments such as basketball and football stadiums.

It's worth noticing that the encoder from our model is initially trained in an unsupervised learning stage using large-scale, freely available unlabeled audio data. In the second stage, this encoder is applied, and a fully connected layer is used to adapt the encoder’s representations for specific domain tasks, such as vibration classification.

By leveraging the abundance of unlabeled audio data, this approach enhances the model's accuracy and robustness in classifying vibrations, even with limited labeled data. It is particularly effective in scenarios where collecting labeled vibration data is costly or challenging, thereby reducing the need for extensive labeled datasets. The details of our implementation are illustrated in the following.

\subsection{Unsupervised pre-training using audio} \label{sec:unsupervised_training_process}


In the first stage, the proposed method uses audio data to learn the wave behaviors, leveraging publicly available datasets and thus reducing the dependency on specific domain vibration data collection. To be more specific, the audio data first transforms into spectrograms and then is split into patches for the following recovery procedure. In this stage, the model learns the wave behaviors from audio and applies these representations to vibration for learning crowd behaviors in the next stage.

To model the modality difference between audio and vibration, the proposed method also down-sampling audio data to meet the low frequency of vibration data. On the other hand, the masking size is also decreased for the lower frequency of vibration data to make the learning procedure more reasonable.

\begin{figure}[h]
    \centering
    \includegraphics[width=1.0\linewidth]{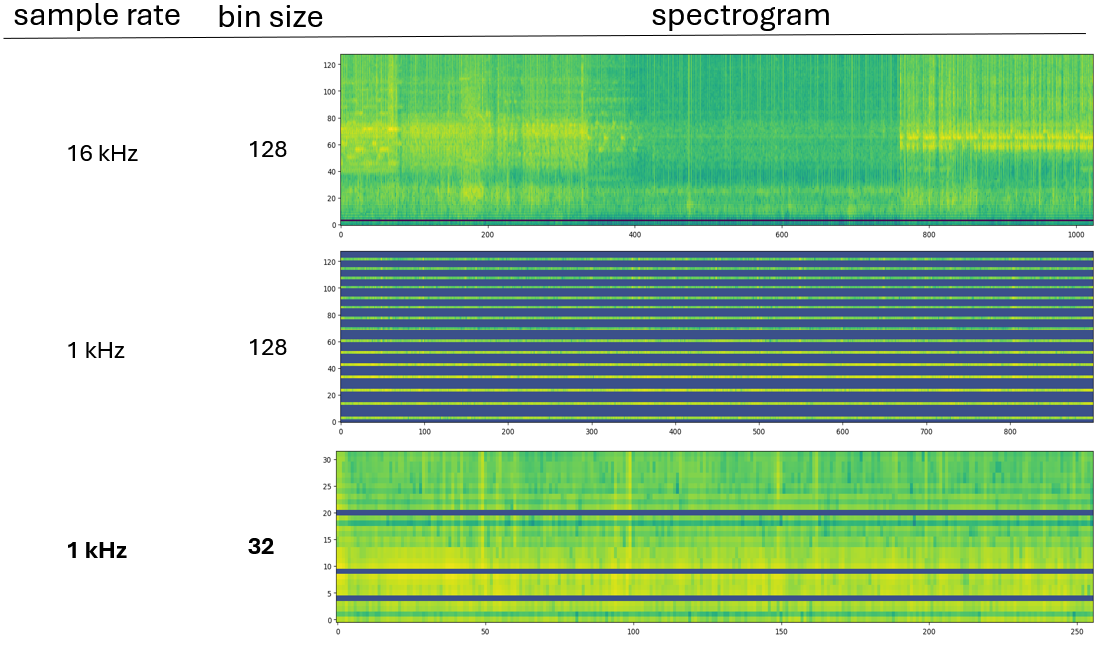}
    \caption{The illustration of audio spectrogram transformation parameter settings. Audio signals are down-sampled from 16 kHz to 1 kHz and converted from 128-bin to 32-bin Mel-spectrograms to prevent null values that can arise from using high bin sizes with low vibration sampling rates.}
    \label{fig:audio_spectrogram}
\end{figure}

\begin{figure}[h]
    \centering
    \includegraphics[width=1.0\linewidth]{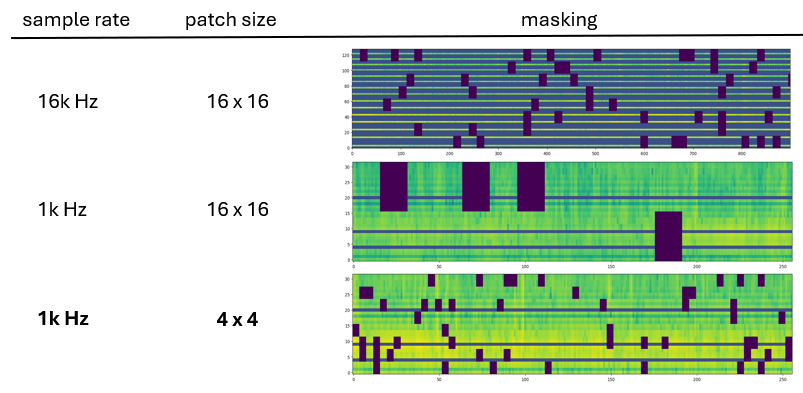}
    \caption{The illustration of masking parameter settings. We used a sample rate of 1 kHz and a patch size of 4×4 to avoid null values in the vibration spectrum, which can occur with lower sampling rates compared to audio signals.}
    \label{fig:masking}
\end{figure}

\textbf{Masked audio.} Initially, we transform our audio signals into Mel-spectrograms. To accommodate the low sampling rate of vibration signals, we down-sample the audio dataset to 1 kHz and utilize 32 bins for the Mel-spectrum transformation to avoid null values, as shown in Figure \ref{fig:audio_spectrogram}. Subsequently, we apply masking to our audio spectrograms for spectrogram recovery, which helps our models learn better representations of these spectrograms. Due to the lower sampling rate of vibration signals, we use a smaller patch size of 4×4 to maintain a reasonable masking ratio after down-sampling, as shown in Figure \ref{fig:masking}. The details of the training loss with different parameter settings are shown in Table \ref{table:model_parameter_ablaton}, and the details on the masking strategies and data augmentation can be found in \cite{huang2022}.

\textbf{Encoder.} After masking our audio spectrograms, we split them into patches and use our encoder to generate embedded vectors. Specifically, we employ standard Transformer blocks, using the 12-layer ViT-Base (ViT-B) model \cite{vaswani2017attention} for encoding tasks. Building on the successful results from \cite{huang2022}, we process only non-masked patches to reduce computational overhead.

\textbf{Decoder.} Following the encoder, a decoder is adopted to recover our masked audio spectrograms. Specifically, the decoder employs local attention using standard Transformer blocks, and we use a time-frequency positional encoder to aid in the recovery process. The implementation details can be found in \cite{huang2022}.

\textbf{Objective for unsupervised learning.} We use mean squared error (MSE) between the original spectrogram and the recovered spectrogram as our objective. Masked Autoencoders utilize the reconstruction process to learn better representations for images. Similarly, for spectrograms, we also employ MSE as our objective, enabling the models to learn the recovery process effectively.

\subsection{Supervised fine-tuning for specific domain}
In this paper, we focus on classification tasks as our specific downstream application, whereas the approach proposed in \cite{dong2023} involves different settings. Our approach involves single-class classification for each 1-minute vibration spectrum. We utilize the encoder from the unsupervised pre-training stage and add a fully connected layer to handle our vibration classification task. The decoder pre-trained on audio data is not used in this stage. Details on converting time-sequence labeling to single-class labels are provided in Section \ref{sec:experiment_setting}.

\textbf{Vibration.} We first convert our 1 kHz vibration signal into spectrograms with 32 frequency bins and a length of 128 to reduce the risk of null values that can arise from excessively large bin sizes. The parameter settings for the spectrogram transformation are illustrated in Figure \ref{fig:audio_spectrogram}. These labeled vibration spectrograms are then used for supervised learning to perform single-class classification tasks. Table \ref{table:model_parameter_ablaton} details the selection of sampling rate and bin size, along with a comparison of the resulting training losses.

\textbf{Encoder.} We applied the encoder pre-trained through unsupervised learning, which has learned a more effective representation space for spectrograms. Similar to the unsupervised pre-training stage, we split the audio spectrograms into patches and use the encoder to generate their embedded vectors. At this stage, we leverage the knowledge gained from the out-of-domain dataset to benefit our vibration-based models.

\textbf{Fully connected layer.} After we obtained the embedded vectors from the audio-trained encoder, a fully connected layer is adopted to learn our vibration down steam classification task. Inspired by \cite{huang2022, oord2018representation, koutini2021efficient}, we also remove a portion of patches from the spectrograms and directly connect the encoder to the fully connected layer. This approach reduces computational overhead during fine-tuning.

\textbf{Objective for supervised fune-tuning} We use cross-entropy as our objective for supervised leaning of vibration tasks. In our experiments, we demonstrate a 7-class classification task to evaluate our vibration model's performance. The details of experiment settings are shown in \ref{sec:experiment_setting}.

\section{Wave Behavior Analysis: Similarity and Diversity} \label{pretrain_modality}
In this section, we are trying to show audio modality is the most suitable modality for vibration-based models, compared to other modalities such as images and video.
Thus, we also design two indicators to measure the physical properties between vibration and other modalities.
In the rest of the subsections, we first introduce two indicators, \textit{Similarity} and \textit{Diversity}, to efficiently identify the preferred pre-training modality, and then use them to assess the effectiveness of various modalities on improving the performance of our vibration-based models. We found that audio data is the most suitable modality for vibration pre-training given its high similarity to the vibration data and high diversity among the existing datasets. Furthermore, we compare various specific types of audio data and select the dataset that will effectively enhance our models. 




\subsection{Indicators for dataset modality selection}
\begin{figure*}
    \centering
    \includegraphics[width=\linewidth]{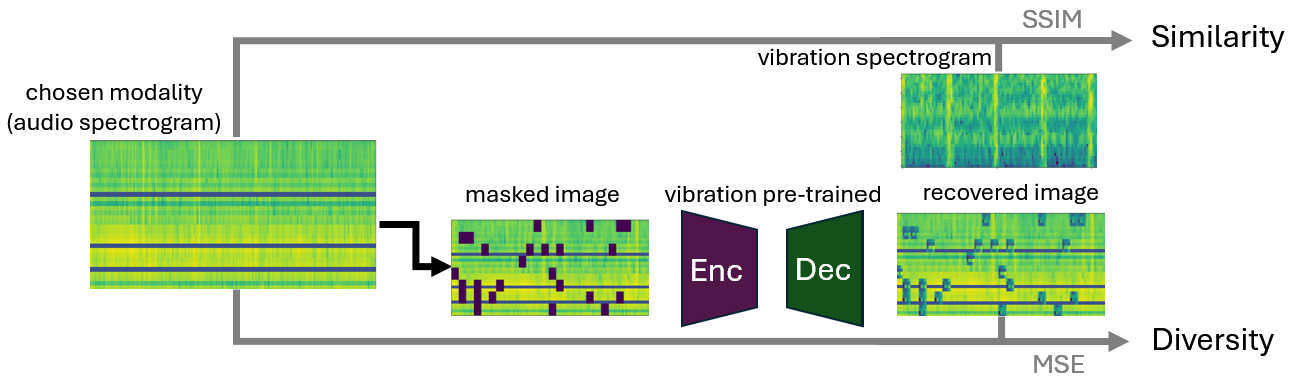}
    \caption{Illustration of \textit{Similarity} and \textit{Diversity} metrics. We use the Structural \textit{Similarity} Index Measure  (SSIM) to measure \textit{Similarity} and Mean Squared Error (MSE) to assess \textit{Diversity}.}
    \label{fig:sim_div}
\end{figure*}

In this subsection, we introduce \textit{Similarity} and \textit{Diversity} indicators to efficiently identify the preferred modality to enhance vibration-based crowd monitoring.
\textit{Similarity} refers to the shared properties between the vibration signal and the selected modality. 
\textit{Diversity} represents the additional information the vibration model gains from the chosen modality. 
The indicators are defined based on feature comparison using spectrograms. Figure \ref{fig:sim_div} illustrates the calculation of these two indicators.

\textbf{\textit{Similarity} – the resemblance between pre-training dataset modality and vibration data.} 
To measure the resemblance between a vibration spectrum dataset and a chosen spectrum or image dataset, we use the Structural \textit{Similarity} Index Measure (SSIM) \cite{wang2017}.
SSIM is commonly used for assessing structural \textit{Similarity} in images. 
A higher SSIM value indicates greater \textit{Similarity} between the chosen modality and the vibration spectrograms.

Specifically, we first convert the vibration signals into spectrograms. 
We then use SSIM to evaluate the pair-wise structural \textit{Similarity} between these spectrograms and the spectrograms of the chosen modality. 
Unlike Mean Square Error (MSE), which is used in Mask Autoencoder reconstruction \cite{he2022}, SSIM is more suitable for estimating spectrograms due to its ability to capture luminance and contrast. 
These factors are crucial for capturing the structural information and context of vibration signals in relation to other modalities \cite{sheng2019high, arun2019automated}. 

The \textit{Similarity} is calculated as follows:
\begin{equation}
\text{\textit{Similarity}}(x, y) = \frac{(2 \mu_x \mu_y + C_1) (2 \sigma_{xy} + C_2)}{(\mu_x^2 + \mu_y^2 + C_1) (\sigma_x^2 + \sigma_y^2 + C_2)}
\end{equation}
where \( \mu_x \) and \( \mu_y \) represent the average pixel values of images \( x \) and \( y \), respectively. \( \sigma_x^2 \) and \( \sigma_y^2 \) denote the variances of images \( x \) and \( y \), while \( \sigma_{xy} \) indicates the covariance between the two images. The constants \( C_1 \) and \( C_2 \) are included to stabilize the division and prevent instabilities when the denominator is close to zero.

\textbf{\textit{Diversity} – the information from the modality that enhances vibration models.} 
To estimate \textit{Diversity}, we use the reconstruction ability of a vibration pre-trained model as our indicator, given our focus on accuracy-oriented measurement. 
Higher \textit{Diversity} indicates that the chosen modality provides more information to the vibration-based model.

Specifically, we first pre-train a reconstruction model using our chosen modality, as described in Section \ref{sec:training_process}. 
Next, we use this reconstruction model to recover spectrograms transferred from the chosen modality. 
We then calculate the Mean Squared Error (MSE) between the recovered and original spectrograms to quantify \textit{Diversity}. 
MSE is employed because it aligns with the objective function used in our method during the pre-training phase, which will be introduced in Section \ref{sec:unsupervised_training_process}. If a model cannot recover a spectrogram from the chosen modality, the spectrogram is likely more diverse compared to the modality's pre-training dataset.

\textit{Diversity} is calculated as follows:
\begin{equation}
\text{\textit{Diversity}} = \frac{1}{mn} \sum_{i=1}^{m} \sum_{j=1}^{n} [I(i,j) - K(i,j)]^2
\end{equation}
where \( I \) and \( K \) are the original and distorted images, respectively, and \( m \) and \( n \) are the dimensions of the images.

\subsection{Identify our pre-trained modality}
In this section, we use the previously proposed indicators to identify which of our chosen modalities most effectively enhances the performance of vibration-based models, thus avoiding the inefficiency of testing each dataset across all modalities. 



\begin{figure}[h]
    \centering
    \includegraphics[width=0.9\linewidth]{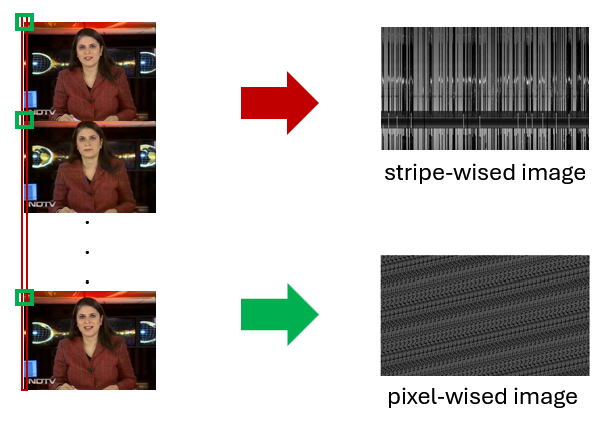}
    \caption{An illustration of our strip-wise and pixel-wise images is provided. We convert our video into time-series images by using the same column (strip-wise) and same pixel (pixel-wise) of each frame.}
    \label{fig:video_images}
\end{figure}

With the rapid advancement of smart sensing technology, numerous intelligent applications in sound, image, and video have emerged, leading to an expansion of publicly available datasets for these modalities. 
Based on these findings, we first utilized audio, image, pixel-wise video, strip-wise video, and vibration modalities to pre-train our models. 
Subsequently, we fine-tuned the models for our downstream vibration classification task to compare their accuracy and analyze their \textit{Similarity} and \textit{Diversity}. 
For the audio modality, we use the YouTube8M dataset \cite{abu2016youtube} as our pre-training dataset, which includes 350,000 hours of video and 3,862 classes of audio. For the image modality, we use the ImageNet-1k dataset \cite{dosovitskiy2020image}, which contains 1,281,167 training images across 1,000 classes. For the video modality, we also use the YouTube8M dataset \cite{abu2016youtube}. We convert the videos into two types of images: strip-wise and pixel-wise. Strip-wise video extracts the same column from each video frame, while pixel-wise video extracts the same pixel from each video frame. The illustrations of the strip-wise and pixel-wise video representations are shown in Figure \ref{fig:video_images}. For the vibration modality, we use the vibration signal data from Stadium 2 as provided by \cite{dong2023}, which includes 21,840 sections of 1-minute vibration recordings.

After comparing various sensing modalities, we chose the audio modality because of its high \textit{Similarity} and high \textit{Diversity}. Figure \ref{fig:audio_pretrain} shows examples of both vibration and audio spectrograms, highlighting the many similar patterns between them. It is noteworthy that although video modalities have higher \textit{Diversity}, their lower \textit{Similarity} results in poorer performance compared to the audio pre-trained model.
The details of the experimental are shown in \ref{sec:modality_comparison} and the results are shown in Table \ref{table:all_pretrain_acc}. 
Based on these findings, we will focus on the audio modality as the primary research topic in the subsequent experiments.

\subsection{Genre comparison within audio data}
\begin{figure}[!h]
    \centering
    \includegraphics[width=1.0\linewidth]{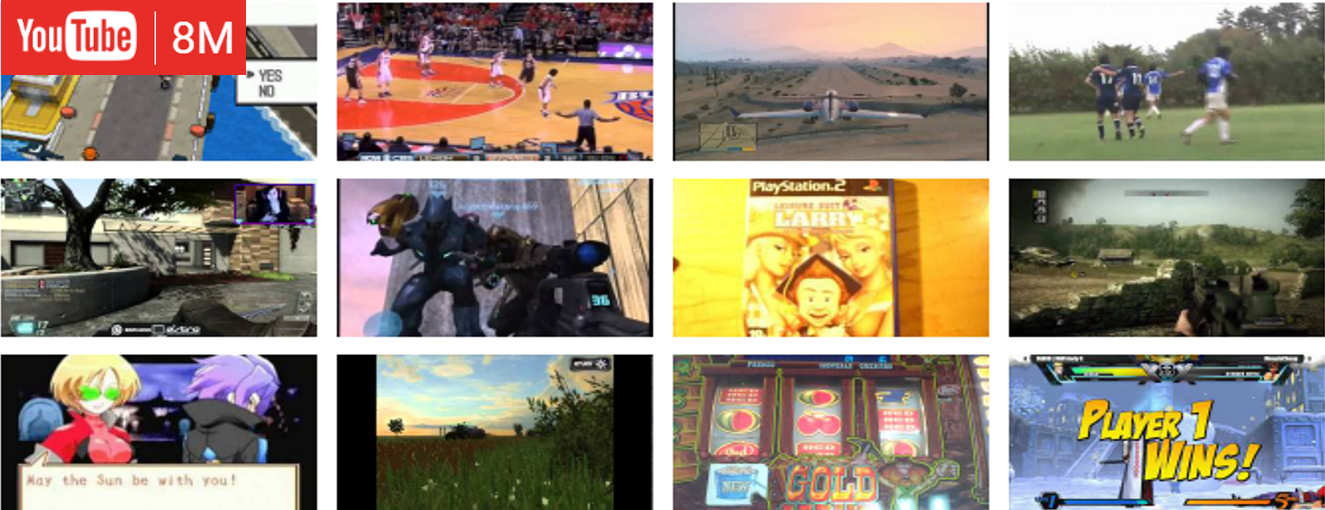}
    \caption{Examples of videos from the YouTube8M dataset. This dataset includes 1,000 self-defined video classes.}
    \label{fig:youtube_genre}
\end{figure}

In addition to comparing the performance of different modalities, we also use the audio genres to further explore which types of audio data can most effectively improve vibration-based models. 
The genre classifications are defined by \cite{abu2016youtube}, as shown in Figure \ref{fig:youtube_genre}.
Given the substantial data available for music and speech, as shown in Figure \ref{fig:top10_music_genre}, we first focus on these two genres. 

The experiment from \ref{sec:genre_comparison} shows that models pre-trained with music audio outperform those pre-trained with speech audio. 
Additionally, the results indicate that subsets with higher \textit{Similarity} achieve better performance, which is consistent with previous conclusions. 
The details of the experimental results are shown in Table \ref{table:all_pretrain_acc}.

\textbf{\textit{Similarity} and \textit{Diversity} analysis.} 
In addition to grouping based on different modalities and audio genres, we use the two indicators proposed in this paper to categorize our audio dataset. This allows us to further examine the relationship between these indicators and the accuracy of the models.
We conduct experiments under two settings: 1) sorting by \textit{Similarity} and \textit{Diversity} independently, and 2) sorting first by \textit{Similarity} and then by \textit{Diversity}. 
These experiments are conducted using the YouTube8M audio dataset, which shows the best performance according to Table \ref{table:all_pretrain_acc}. The details of experiments are shown in Section \ref{sec:sd_comparison}.

The experiment of \ref{sec:sd_comparison} indicates that intermediate values of \textit{Similarity} or \textit{Diversity} tend to yield better performance in audio dataset. 
It is important to note that since we divide the dataset into three groups, the amount of data in each group is reduced to one-third, which can lead to poorer performance comparing with we use all the audio data to do pre-training. 
The experimental results are shown in Table \ref{table:sortting_3}.

In addition to analyzing the performance based on each indicator individually, we also conduct experiments that combine the values of both indicators. 
We first sort the dataset by \textit{Similarity}, which, as shown in Table \ref{table:all_pretrain_acc}, is the more significant factor comparing to \textit{Diversity}. 
After that, we then further sort these subgroups by \textit{Diversity}. 
The experimental results from \ref{table:sortting_9} indicate that intermediate or higher values for both \textit{Similarity} and \textit{Diversity} tend to yield slightly better performance compared to other configurations, as illustrated in Table \ref{table:sortting_3}. 
It is also important to note that a smaller data amount may lead to poorer performance.

\section{Real-world during-game stadium evaluation}
In this section, we outline our experimental setup, including sensor deployment, training configurations, and label definitions for classification. We then detail our approach to parameter and modality selection based on this setup. Finally, we assess the performance of our proposed method through evaluations conducted at two distinct ball game stadiums.

\subsection{Experiment settings} \label{sec:experiment_setting}
\begin{figure}
    \centering
    \includegraphics[width=\linewidth]{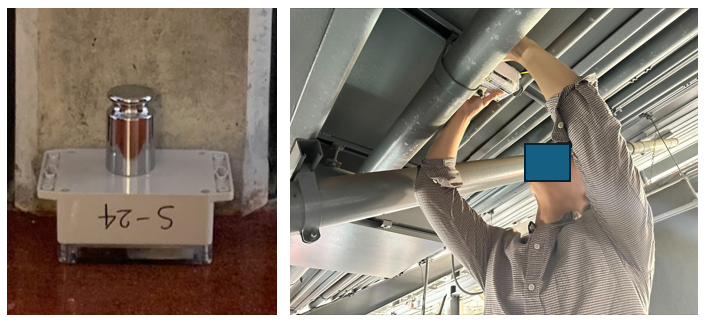}
    \caption{Sensor installation in Stadium 1. The vibration device is shown on the left, and the sensor deployment demonstration is shown on the right. Sensors are strategically placed beneath the cement in seating areas to enhance connectivity.}
    \label{fig:sensor}
\end{figure}

For our experimental setup, we deploy six geophones as our vibration sensors in each game venue to monitor crowd dynamics from the floor. We set the sampling frequency to 500 Hz in Stadium 1 and 1000 Hz in Stadium 2. These frequencies were chosen to balance temporal resolution with data transmission efficiency, tailored to the varying crowd sizes and dynamics at each stadium. All sensors are connected via wireless routers to ensure effective data transmission. In Stadium 1, which hosts a large audience of over 100,000, sensors are installed beneath the cement in seating areas to improve connectivity. Figure \ref{fig:sensor} illustrates the deployment and sensor setup in Stadium 1. Further details on the deployment for each stadium are provided in Sections \ref{sec:umich_experiemnt} and \ref{sec:stanford_experiemnt}.

For our training setup, we utilize an NVIDIA RTX A6000 with 48GB of memory, with each model undergoing 120 hours of training. We use vibration, audio, image, and video data as our pre-training modalities.
\begin{itemize}
    \item \textbf{Vibration:} We use data from \cite{dong2023}, which includes 21,840 sections of 1-minute vibration recordings. Further details are available in \cite{dong2023}.
    \item \textbf{Audio:} For audio pre-training, we employ the YouTube8M dataset \cite{abu2016youtube} for unsupervised learning. Due to the removal of some videos, we used the remaining videos to generate 19,420 1-minute audio signals. Additional information can be found in \cite{abu2016youtube}.
    \item \textbf{Image:} We use a subset of ImageNet \cite{5206848} consisting of 20,000 images to match the data volume of other modalities.
    \item \textbf{Video:} We also use a subset of YouTube8M for video data. To maintain the time series properties and convert video data to 2D images, we apply two different settings:
    \begin{itemize}
        \item \textbf{Stripe-wise:} Extracts the same column from each frame.
        \item \textbf{Pixel-wise:} Extracts the same pixel from each frame.
    \end{itemize}For each setting, we generate 20,000 images. An illustration of how we generate video data is shown in Figure \ref{fig:video_images}.
\end{itemize}

\begin{figure}[h]
    \centering
    \includegraphics[width=\linewidth]{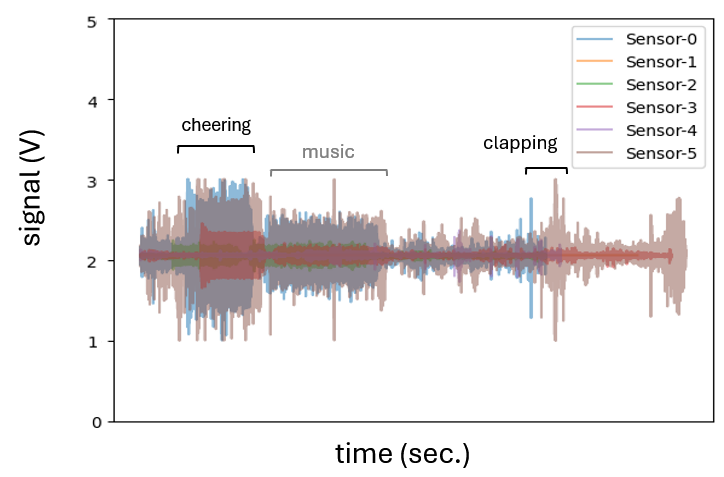}
    \caption{In our classification model, more intense behaviors such as cheering take precedence over milder ones like clapping.}
    \label{fig:classification_define1}
\end{figure}

\begin{figure}[h]
    \centering
    \includegraphics[width=1.0\linewidth]{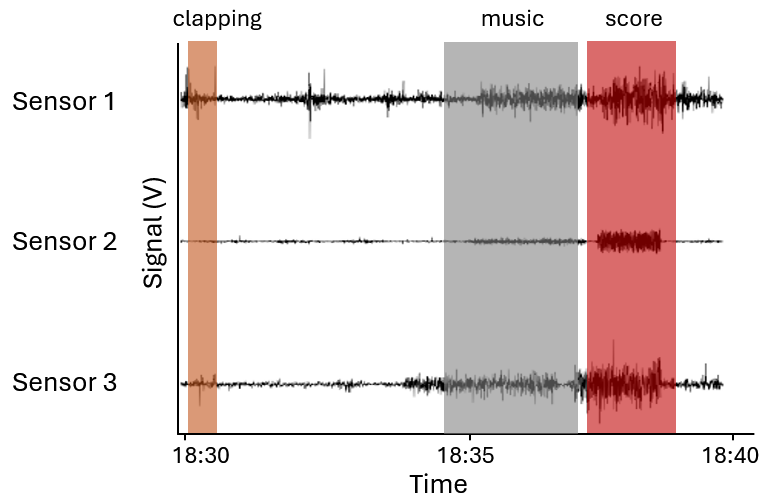}
    \caption{Despite multiple sensors in different locations, we captured consistent behavior across multiple locations.  }
    \label{fig:classification_define2}
\end{figure}

The proposed method is a 7-class classification model for every single minute. Since each minute will have a single class, we assume that more prominent behaviors (such as booing and stomping) will overshadow less noticeable behaviors (such as moving and quiet) in our experiments. Therefore, our behavior classes are prioritized as follows: booing, stomping, cheering, clapping, moving, active, and quiet. For instance, if there are both cheering and clapping in a single minute, we will classify this behavior as cheering as shown in Figure \ref{fig:classification_define1}. Additionally, since we use game play-by-play records (e.g., home team scoring often results in stomping and away team scoring tends to be quieter) to label the data, we ensure consistency in labeling by recording only one behavior. We select the most intense sensor from all locations based on our defined priority order, as shown in Figure \ref{fig:classification_define2}.

\subsection{Experiments for parameter and modality selection}
In this section, we set Stadium 2 on February 17th as our default downstream task because this dataset provides the largest amount of data. We first conduct experiments to optimize the parameter settings for unsupervised learning. Next, we evaluate which of the publicly available modalities most effectively enhances our vibration-based models. After identifying that the audio modality outperforms the others, we investigate which music genres are most effective for further improving our vibration models. Finally, we use our proposed indicators to assess the relationship between model accuracy and these indicators.

\subsubsection{Parameter settings} \label{sec:modality_comparison}
To apply unsupervised learning methods to the vibration-based model, we perform an ablation study to determine optimal parameters for spectrogram transformation and masking procedures. We evaluate the performance of the unsupervised learning approach using training loss, which reflects the reconstruction accuracy based on mean squared error. Details of the reconstruction procedures are provided in Section \ref{sec:unsupervised_training_process}. We use the vibration dataset as our pre-training modality. The results indicate that a spectrogram size of 256$\times$32 (time$\times$bin) and a patch size of 4$\times$4 (unit$\times$unit) yield the best performance, as shown in Table \ref{table:model_parameter_ablaton}.

\begin{table}[h]
\begin{tabular}{cc|c}
spectrum   size & patch size   & training loss   \\ \hline
1024$\times$128        & 16$\times$16        & 0.3838          \\
1024$\times$32         & 16$\times$16        & 0.1747          \\
256$\times$32          & 16$\times$16        & 0.1773          \\
\textbf{256x32} & \textbf{4x4} & \textbf{0.1214}
\end{tabular}
\caption{The training loss in different kinds of spectrum size and patch size settings. }
\label{table:model_parameter_ablaton}
\end{table}

\subsubsection{Modality comparison} \label{sec:modality_comparison}
After determining the optimal model parameters and masking settings, we use these settings to identify which modalities can most effectively improve our vibration models. We first pre-train our models with the pre-training modalities mentioned in Section \ref{sec:experiment_setting}, then fine-tune them using data from Stadium 2 on February 17th. The experiments show that the audio modality has the highest \textit{Similarity} to vibration and achieves the highest accuracy among all modalities. Although the vibration modality itself has the highest \textit{Similarity}, its lower \textit{Diversity} leads to poorer performance. Conversely, while video modalities exhibit higher \textit{Diversity}, their lower \textit{Similarity} results in suboptimal performance. The experimental results are shown in Table. \ref{table:all_pretrain_acc}.

\begin{table*}[ht]
\centering
\begin{tabular}{c|c|c|ccc|c|cc}
         & \begin{tabular}[c]{@{}c@{}}train from \\ scratch\end{tabular} & vibration & \begin{tabular}[c]{@{}c@{}}all\\ audio\end{tabular} & \begin{tabular}[c]{@{}c@{}}music\\ audio\end{tabular} & \begin{tabular}[c]{@{}c@{}}speech\\ audio\end{tabular} & image & \begin{tabular}[c]{@{}c@{}}point-wise\\ video\end{tabular} & \begin{tabular}[c]{@{}c@{}}stripe-wise\\ video\end{tabular} \\ \hline
sim.     &                                                                 & 0.35      & \textbf{0.24}                                       & 0.21                                                  & 0.19                                                   & 0.11  & 0.13                                                       & 0.16                                                        \\
div.     &                                                                 & 0.04      & \textbf{0.22}                                       & 0.14                                                  & 0.13                                                   & 0.21  & 0.31                                                       & 0.29                                                        \\ \hline
acc. \% & 27.39                                                           & 39.97     & \textbf{87.51}                                      & 30.56                                                 & 24.73                                                  & 28.89 &  25.36                                                     & 41.54                                                      
\end{tabular}

\caption{Overview of our empirical experiments with vibration-based models, comparing different pre-training modalities and our proposed indicators. As shown in the table, the model pre-trained on vibration signals exhibits the highest \textit{Similarity}, while pre-training with audio signals outperforms other modalities.}
\label{table:all_pretrain_acc}
\end{table*}

\subsubsection{Genre comparison} \label{sec:genre_comparison}
\begin{figure}[!h]
    \centering
    \includegraphics[width=1.0\linewidth]{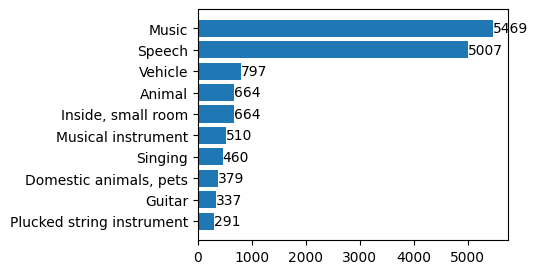}
    \caption{Analysis of data volume using audio from remaining YouTube8M. The results indicate that music and speech audio categories account for over five times the amount of data compared to other genres.}
    \label{fig:top10_music_genre}
\end{figure}

In previous experiments, we identified that the audio modality outperforms other chosen publicly available modalities. In this experiment, we further explore which type of audio genre most efficiently improves vibration models. The audio genres were defined by \cite{abu2016youtube}, which follows expert labeling for genre classification. Due to the significant differences in data amounts, as shown in Figure \ref{fig:top10_music_genre}, we selected music and speech as our testing genres. The experimental results show that the music genre outperforms the speech genre. Additionally, the values of \textit{Similarity} and \textit{Diversity} for music are higher than those for speech. Detailed experimental results are presented in Table \ref{table:all_pretrain_acc}.

\subsubsection{\textit{Similarity} and \textit{Diversity} comparison} \label{sec:sd_comparison}
To further understand the relationship between our indicators (\textit{Similarity} and \textit{Diversity}) and the model's performance, we conducted additional experiments. We used the audio modality as our pre-training data and grouped the data in two different settings: 
1. 3 groups sorted by either \textit{Similarity} or \textit{Diversity}.
2. 9 groups sorted first by \textit{Similarity} and then by \textit{Diversity}.
Given the results from Section \ref{sec:modality_comparison}, which indicate that \textit{Similarity} is more important than \textit{Diversity}, we primarily evaluate the order sorted by \textit{Similarity} and then by \textit{Diversity}. Table \ref{table:sortting_3} shows the experimental results for the three-group setting. These experiments reveal that intermediate values of \textit{Similarity} or \textit{Diversity} outperform the other values. Table \ref{table:sortting_9} presents the results for the nine-group setting, indicating that intermediate or higher values of both \textit{Similarity} and \textit{Diversity} outperform the others. It is important to note that dividing the dataset into three groups reduces the amount of data in each group to one-third, which can lead to poorer performance compared to using all the audio data for pre-training.

\begin{table}[h]
\begin{tabular}{l|lll}
 & low    & mid             & high   \\ \hline
sim.       & 23.03  & \textbf{32.04}  & 22.04  \\
div.       & 25.57  & \textbf{38.19}  & 28.07  \\ \cline{1-4} 
all       & \multicolumn{3}{c}{\textbf{87.5}}
\end{tabular}
\caption{Experimental results for two 3-group subsets of the audio dataset. YouTube8M dataset is divided into 3 different groups based on our defined \textit{Similarity} or \textit{Diversity}.}
\label{table:sortting_3}
\end{table}

\begin{table}[h]
\begin{tabular}{l|lll}
sim. \symbol{92} div.                                          & low            & mid            & high           \\ \hline
\multicolumn{1}{l|}{low}  & 20.50 & 20.08          & 20.27          \\
\multicolumn{1}{l|}{mid} & 20.40 & \textbf{21.40} & 19.85          \\
\multicolumn{1}{l|}{high} & 19.18 & 19.35          & \textbf{21.19}
\end{tabular}
\caption{Experimental results for a 9-group subset of the audio dataset. YouTube8M dataset is divided into 9 different groups based on our defined \textit{Similarity} and \textit{Diversity}.}
\label{table:sortting_9}
\end{table}

\subsection{Experiments in real-game deployments}
\subsubsection{Football game at Stadium 1} \label{sec:umich_experiemnt}
\begin{figure}[h]
    \centering
    \includegraphics[width=1.0\linewidth]{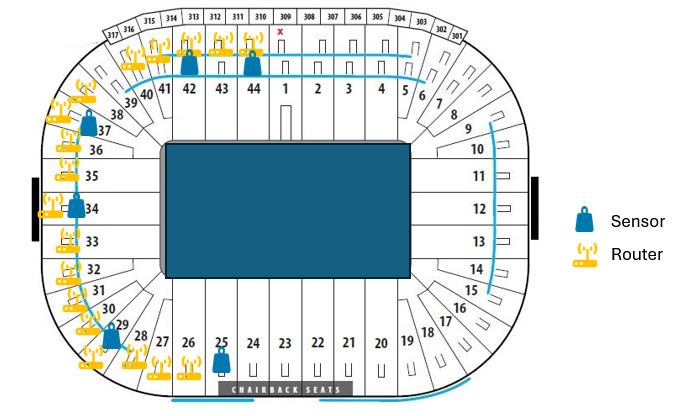}
    \caption{An illustration of the sensor and router deployments at Stadium 1. The yellow icons represent the routers, and the blue icons indicate our sensors.}
    \label{fig:umich_deploymtny}
\end{figure}

In this experiment, we evaluate the performance of pre-training models across different modalities. Our deployment of routers and sensors is illustrated in Figure \ref{fig:umich_deploymtny}. Given that Stadium 1 has an official capacity of 100,000, maintaining reliable sensor connectivity is challenging. To optimize our limited resources, we deploy sensors in only half of the stadium, which enhances the reliability of router connections. The experiments are conducted on November 4th and 30th, during the Michigan vs. Purdue and Ohio State games, respectively.

\begin{figure}[h]
    \centering
    \includegraphics[width=0.7\linewidth]{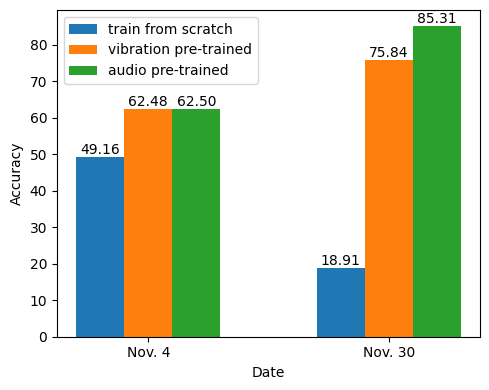}
    \caption{The experimental results of comparing different modalities at the Stadium 1.}
    \label{fig:umich_plot}
\end{figure}

The experimental results demonstrate that pre-training with audio data significantly outperforms other methods. Models that utilized pre-training generally performed better compared to those trained from scratch. Notably, the audio pre-training model achieved an accuracy that was 4.5 times higher on November 30th compared to models trained from scratch. This significant improvement underscores the effectiveness of audio pre-training in enhancing model performance. Details of the evaluation are illustrated in Figure \ref{fig:umich_plot}.

\subsubsection{Basketball game at Stadium 2} \label{sec:stanford_experiemnt}

\begin{figure}[h]
    \centering
    \includegraphics[width=1.0\linewidth]{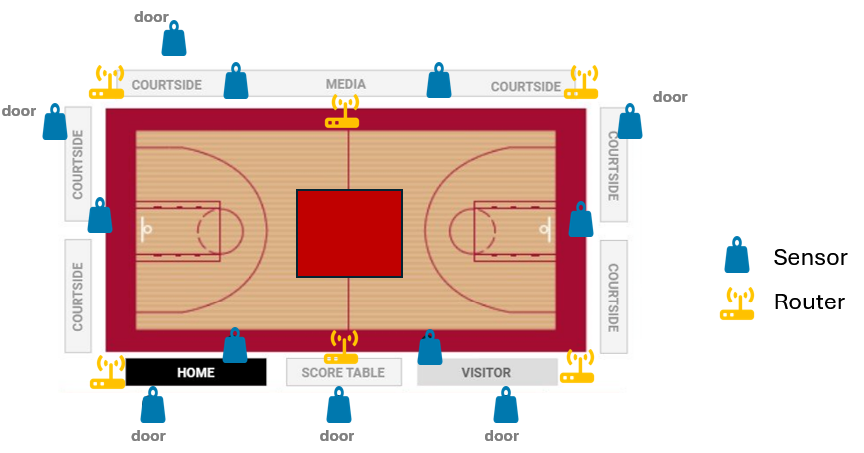}
    \caption{An illustration of the sensor and router deployments at Stadium 2. The yellow icons represent the routers, and the blue icons indicate our sensors.}
    \label{fig:stanford_deployment}
\end{figure}

In our Stadium 2 crowd monitoring experiments, we use the dataset from \cite{dong2023} but in our classification setting. We transfer the time-series prediction problem into single vibration section perdition problem in the priority mentioned in \ref{sec:experiment_setting}. Figure \ref{fig:stanford_deployment} shows the deployments of vibration sensors in the stadium and further deployment detail can be reference in \cite{dong2023}.

In Stadium 2, the experimental results are consistent with those from Stadium 1. The results also show that audio pre-trained models outperform other methods. Specifically, the accuracy of pre-trained models from February 17th and 20th is 4.2 times and 3.3 times higher, respectively, than models trained from scratch. These experimental results indicate that pre-training with audio datasets can significantly improve the performance of vibration-based models. Details of the evaluation are illustrated in Figure \ref{fig:stanford_deployment}.

\begin{figure}
    \centering
    \includegraphics[width=0.7\linewidth]{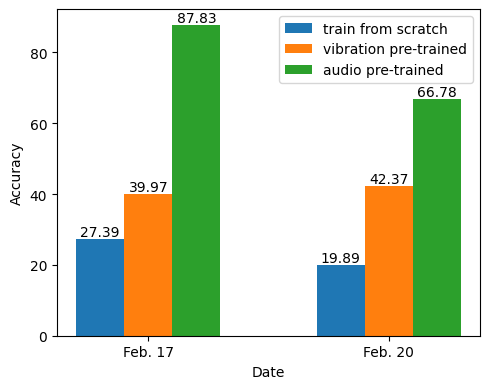}
    \caption{The experimental results of comparing different modalities at the Stadium 2.}
    \label{fig:stanford_plot}
\end{figure}

\section{Conclusion}
In this paper, we introduced the \emph{ViLA} method, a novel approach designed to address the common challenge of limited well-labeled data in sensing applications. By leveraging cross-modality pre-training, \emph{ViLA} utilizes the natural similarities between audio and vibration data to enhance the performance of vibration-based crowd monitoring models. Our method first pre-trains a model using unlabeled audio data, then fine-tunes it with labeled vibration data relevant to the deployment environment.

The introduction of the \textit{Similarity} and \textit{Diversity} indicators has provided a systematic way to evaluate the effectiveness of different modalities in capturing the essential properties of vibration data. These indicators guided our selection of audio data as a pre-training modality, which proved to be particularly effective.

Our extensive real-world experiments in two sports stadiums validate the efficacy of the \emph{ViLA} approach. The results demonstrate that using YouTube8M audio data for pre-training significantly reduces error rates in vibration-based crowd behavior monitoring, achieving up to a 5.8$\times$ reduction compared to models without audio pre-training. This significant improvement highlights the practical benefits of incorporating audio pre-training for accurate and efficient crowd monitoring.

In summary, \emph{ViLA} presents a practical and innovative solution for enhancing vibration-based sensing models, especially in environments where collecting well-labeled data is challenging. Future work will explore extending this approach to other sensing applications and further refining the \textit{\textit{Similarity}} and \textit{\textit{Diversity}} indicators to optimize modality selection.



\bibliographystyle{IEEEtran}
\bibliography{main}

\end{document}